\documentclass[twocolumn,reprint,prb,aps,amsmath,showpacs,floatfix,superscriptaddress]{revtex4-2}

\usepackage{graphicx}
\usepackage{dcolumn}
\usepackage{mathtools}
\usepackage{color}
\usepackage{braket}
\usepackage{hyperref}
\graphicspath{{figures/}}
\usepackage{soul}
\usepackage[dvipsnames]{xcolor}

\usepackage{algorithmicx}

\renewcommand{\S}{S_{11}}
\newcommand{\Qe}{Q_{ext}}
\newcommand{\Qi}{Q_{int}}
\newcommand{\Ql}{Q_{load}}
\newcommand{\fres}{f_0}
\newcommand{\C}{C}
\newcommand{\Cq}{C_q}

\newcommand{\G}{G}
\newcommand{\Qt}{Q_{tot}}
\newcommand{\Vi}{V_{S}}
\newcommand{\Vd}{V_{D}}
\newcommand{\Vb}{V_{B}}
\newcommand{\lnn}{\mathrm{ln}(N)}
\newcommand{\fref}{\fres^{ref}}
\newcommand{\gamt}{\Gamma_t}
\newcommand{\ueV}{$\mu$eV}

\newcommand{\even}{``$E$'' }
\newcommand{\odd}{``$O$'' }

\newcommand{\at}[2][]{#1|_{#2}}

\usepackage[normalem]{ulem}

\newcommand{\qutech}{QuTech and Kavli Institute of Nanoscience, Delft University of Technology, 2600 GA Delft, The Netherlands}
\newcommand{\msdelft}{Microsoft Quantum Lab Delft, Delft University of Technology, 2600 GA Delft, The Netherlands}
\newcommand{\qdev}{Center for Quantum Devices, Niels Bohr Institute, University of Copenhagen, 2100 Copenhagen, Denmark}
\newcommand{\JSI}{Jo\v{z}ef Stefan Institute \& Faculty of Mathematics and Physics, University of Ljubljana, Ljubljana, Slovenia}
\newcommand{\budapest}{Department of Theoretical Physics, Institute of Physics, Budapest University of Technology and Economics, Műegyetem rkp. 3., H-1111 Budapest, Hungary}
\newcommand{\MTA}{MTA-BME Quantum Dynamics and Correlations Research Group, M\H{u}egyetem rkp. 3., H-1111 Budapest, Hungary}
\newcommand{\hyberdad}{School of Physics, University of Hyderabad, Hyderabad, India}

\begin{document}
\title{Quantum capacitance of a superconducting subgap state in an electrostatically floating dot-island}

\author{Filip~K.~Malinowski}
\email{f.k.malinowski@tudelft.nl}
\affiliation{\qutech}

\author{R.~K.~Rupesh}
\affiliation{\hyberdad}

\author{Luka~Pave\v{s}i\'{c}}
\affiliation{\JSI}

\author{Zolt\'an~Guba}
\affiliation{\budapest}

\author{Damaz~de~Jong}
\affiliation{\qutech}

\author{Lin~Han}
\affiliation{\qutech}

\author{Christian~G.~Prosko}
\affiliation{\qutech}

\author{Michael~Chan}
\affiliation{\qutech}

\author{Yu~Liu}
\affiliation{\qdev}

\author{Peter~Krogstrup}
\affiliation{\qdev}

\author{Andr\'as~P\'alyi}
\affiliation{\budapest}
\affiliation{\MTA}

\author{Rok~\v{Z}itko}
\affiliation{\JSI}

\author{Jonne~V.~Koski}
\affiliation{\msdelft}

\date{\today}

\begin{abstract}
We study a hybrid device defined in an InAs nanowire with an epitaxial Al shell that consists of a quantum dot in contact with a superconducting island. The device is electrically floating, prohibiting transport measurements, but providing access to states that would otherwise be highly excited and unstable. Radio-frequency reflectometry with lumped-element resonators couples capacitatively to the quantum dot, and detects the presence of discrete subgap states. We perform a detailed study of the case with no island states, but with quantum-dot-induced subgap states controlled by the tunnel coupling. When the gap to the quasi-continuum of the excited states is small, the capacitance loading the resonator is strongly suppressed by thermal excitations, an effect we dub ``thermal screening''. The resonance frequency shift and changes in the quality factor at charge transitions can be accounted for using a single-level Anderson impurity model.
The established measurement method, as well as the analysis and simulation framework, are applicable to more complex hybrid devices such as Andreev molecules or Kitaev chains.
\end{abstract}

\maketitle


	Andreev bound states~\cite{andreev1966} and Yu-Shiba-Rusinov states~\cite{rusinov1969} are the most familiar types of subgap states (SGSs) observed in Josephson junctions~\cite{pillet2010,bretheau2013}, at atoms on a superconducting surface~\cite{cornils2017} or in semiconducting quantum dots (QDs) coupled to superconductors (SCs)~\cite{lee2014,jellinggaard2016}. SGSs, just like electronic states in QDs, are well localized in space and for odd electron occupancy have a spin which can be manipulated~\cite{chtchelkatchev2003,hays2021,pitavidal2022}. 
The ground state (spin singlet or doublet) depends on the microscopic details~\cite{saldana2021,pavesic2021}. Electrostatic floating of such a device consisting of a dot coupled to a superconductor fixes the total charge, so that the SGS cannot undergo the singlet-doublet phase transition, thereby enabling access to the regimes beyond reach of conventional transport measurements. Furthermore, forcing a fixed charge of the system largely eliminates quasiparticle poisoning that challenges the realization of qubits based on SGSs~\cite{zazunov2003,janvier2015,hays2021,rainis2012}.

\begin{figure}[tb]
	\includegraphics[scale=1]{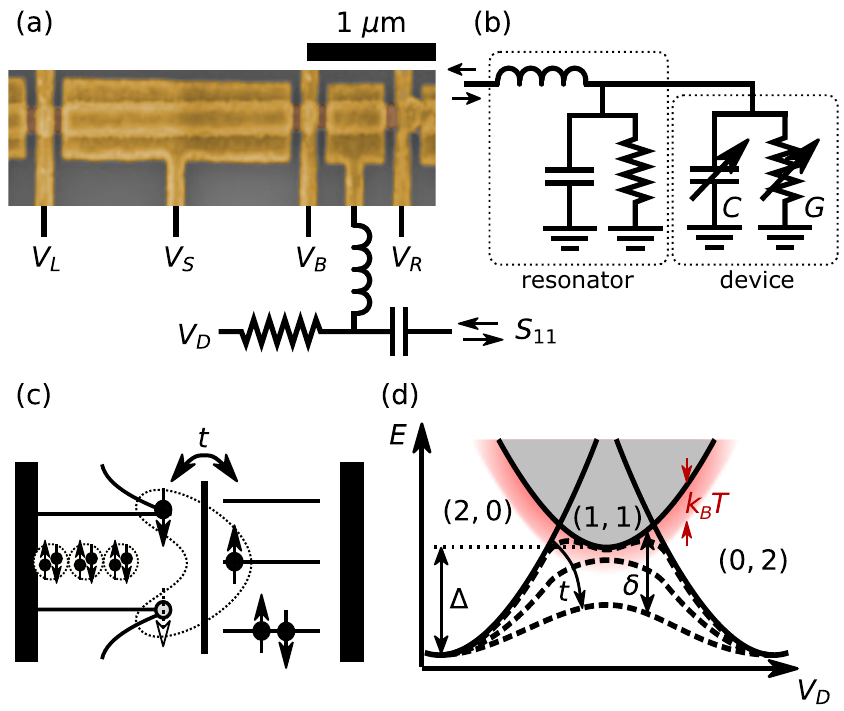}
	\caption{(a) False-colored SEM of the device nominally identical to the one measured, schematically illustrating the resonator circuit with a bias-tee.
			(b) An equivalent resonator circuit. The fixed inductance, capacitance and resistance represent the spiral inductor resonator. The variable capacitance and resistance represent the loading of the resonator due to quantum and tunneling capacitance, and losses in the coupled dot-island system.
			(c) Cartoon illustrating the density of states in a superconducting island, and their coupling to the quantum dot.
			(d) Schematic energy diagram illustrating the evolution of the discrete ground state with increasing tunnel coupling. The gray shaded area represents a quasi-continuum of states, with an unpaired quasiparticle on the superconducting island. The red shaded area represents a range within which the system will likely be thermally excited to the quasi-continuum.
			}
	\label{fig_device}
\end{figure}
	
In this work we study a SGS formed in a QD coupled to a SC island defined in an InAs nanowire. The system is galvanically isolated and the total charge is fixed. We couple the QD capacitively to a radio-frequency resonator~\cite{colless2013} and study the device through its effects on the resonator response. We propose a simple device model that is solved using the density matrix renormalization group (DMRG). We establish that the reactive part of the device response predominantly originates from the charge dispersion, i.e. the charge susceptibility of the instantaneous eigenstates (quantum capacitance). The tunneling capacitance, related to the redistribution of occupancies between the eigenstates during a driving cycle, is significantly smaller, but the associated relaxation process leaves a footprint on the dissipation in the resonator (Sisyphus resistance)~\cite{esterli2019}.


The device under study consists of a nanowire with a two-facet epitaxial Al shell~\cite{krogstrup2015,chang2015} selectively etched away (Fig.~\ref{fig_device}(a)). Wrapped gates are used to electrostatically divide the wire into segments. The left segment, 1.8~$\mu$m long,  is operated as a SC island that is tuned by gate voltage $\Vi$. The right segment, 500~nm long, forms a QD that is tuned by gate voltage $\Vd$. The dot and the island are tunnel coupled with coupling strength $t$ that is controlled by gate voltage $\Vb$ (Fig.~\ref{fig_device}(a,c)). The side barrier gate voltages $V_{L/R}$ are set to large negative values ($< -2$~V) to galvanically disconnect the device and fix its total charge on a timescale of several minutes to days. Additionally, the QD plunger gate is attached to an off-chip spiral inductor resonator~\cite{hornibrook2014} (inductance $L=570$~nH; resonance frequency $\fres \approx 368$~MHz; internal and external quality factors $\Qi \approx 4000$ and $\Qe \approx 285$, respectively). Near an interdot charge transition the electron tunneling between the QD and SC island is enabled, loading the resonator with an additional capacitance $\C$ and conductance $G$, see Fig.~\ref{fig_device}(b) for the effective RLC network model of the setup\footnote{We note that while the enhanced dissipation is quantified in terms of an effective conductance, and in units of $e^2/h$, there is no current flowing through the device. Representing the dissipation as a parallel resistance to ground results in large conductance representing large dissipation. This representation, and the units of $e^2/h$, were chosen to establish a relation to the rf-conductance measurements that can be performed with similar resonators, and to illustrate the magnitude of the effect.}. The resonator loading manifests as a shift of the resonant frequency $\fres$ and a reduction of the internal quality factor $\Qi$ (Appendix~\ref{app_resonator}).


\begin{figure*}[tbh]
	\includegraphics[scale=1]{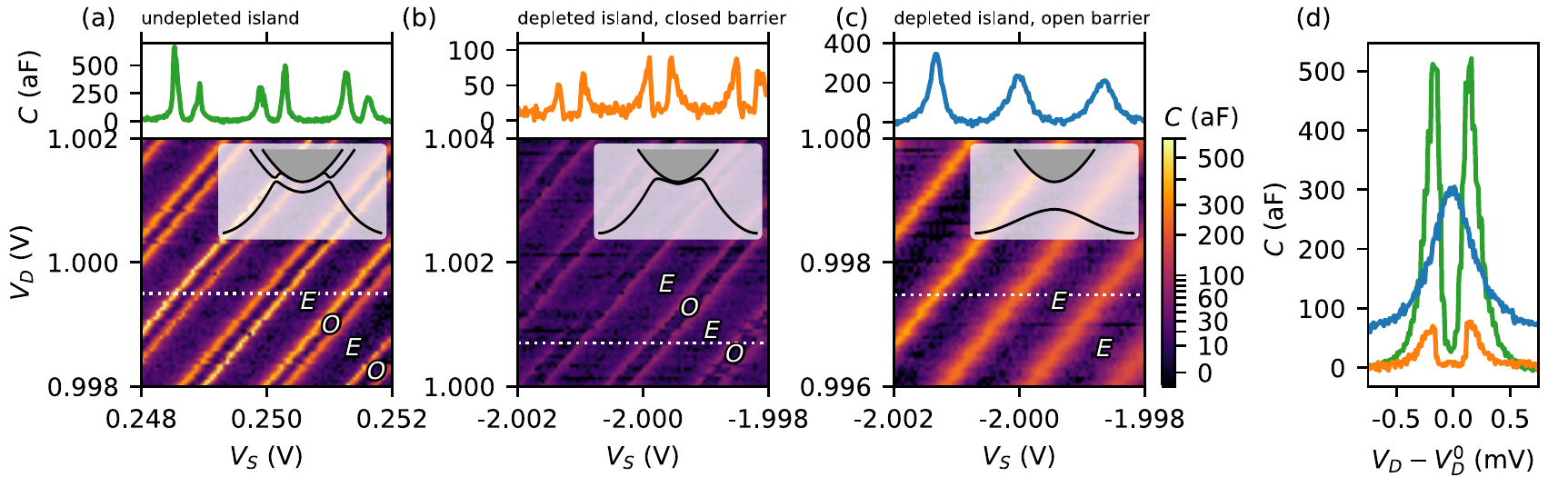}
	\caption{Charge stability diagrams revealed by capacitance $\C$. Panels (a-c) represent three regimes:
	(a) undepleted semiconductor under the Al shell;
	(b) depleted semiconductor under the Al shell and small tunnel coupling;
	(c) depleted semiconductor and large tunnel coupling.
	Shared power-law normalization of color maps enables direct comparison. Top panels show the cuts along the dotted lines, while insets illustrate schematically the energy diagram in each regime (cf.~Fig.~\ref{fig_device}(d)).
	(d) Close-ups on peak patterns in the three regimes; color scheme as in the top panels of (a-c).
	}
	\label{fig_subgap}
\end{figure*}

In order to study the formation of a SGS in the QD we first investigate features indicating whether there are additional discrete SGSs formed in the SC island itself. We start at a moderately positive value of $\Vi \approx 0.25$~V and measure a charge stability diagram by sweeping $\Vi$ and $\Vd$ (Fig.~\ref{fig_subgap}(a)). The measurement reveals a pattern of alternating wider and narrower regions of stable charge (labeled \even and \odd, respectively), separated by charge transitions that are weakly asymmetric with respect to their maximum. Following Ref.~\onlinecite{van2019} we interpret that narrow stability regions \odd correspond to an odd-occupied island, since SC pairing favors even occupancy of the island.
Next, we apply a large negative island gate voltage $\Vi \approx -2$~V. We expect this to deplete the semiconductor wire under the Al shell, eliminating any potential subgap states~\cite{chang2015,vaitiekenas2018,menard2019}. The resulting charge stability diagram shown in Fig.~\ref{fig_subgap}(b) exhibits a similar pattern of narrower and wider regions of charge stability, but the capacitance at charge transitions has a much smaller magnitude and exhibits very strong asymmetry, with a sharp edge on the side of the charge stability regions \odd. Finally, we tune the barrier gate voltage $\Vb$ more positive while keeping $\Vi \approx -2$~V. Fig.~\ref{fig_subgap}(c) shows that the resonance periodicity remains unchanged, however the number of observed interdot charge transitions is halved, the transitions are symmetric, and the added capacitance at the charge transition is increased.

We interpret the three tunings of the device as follows. For $\Vi \approx 0.25$~V, there are one or several discrete subgap states in the island (Fig.~\ref{fig_subgap}(a)). Since the lowest of these states as well as the QD-induced subgap state are well separated from the SC continuum, the charge transitions exhibit many of the same features as those in double QD devices~\cite{van2019}. If the semiconductor is depleted, however, the QD state hybridizes only with the quasicontinuum above the gap.

Fig.~\ref{fig_subgap}(b) represents the case of weak hybridization (small $t$) that allows the QD-induced subgap state to approach the quasicontinuum within $\lesssim kT$. This enables thermal excitation to one of the many states with a single quasiparticle that is decoupled from the QD, suppressing the capacitance (``thermal screening''). Since for an odd-occupied island the ground state approaches the quasicontinuum much more closely (c.f. Fig.~\ref{fig_device}(d)) the thermal screening leads to strong asymmetry of charge transitions.

Fig.~\ref{fig_subgap}(c) corresponds to a strong hybridization (large $t$) in which case the dot-induced discrete ground state becomes well separated from the quasicontinuum. This results in the vanishing of  stability regions \odd, leaving wide charge transitions separating states differing by 2 in the QD occupancy. In the following, we fix $\Vi = -2$~V, thereby eliminating unintended subgap states, and study the transition between the weak and strong hybridization regime in more detail.

\begin{figure*}[tbh]
	\includegraphics[scale=1]{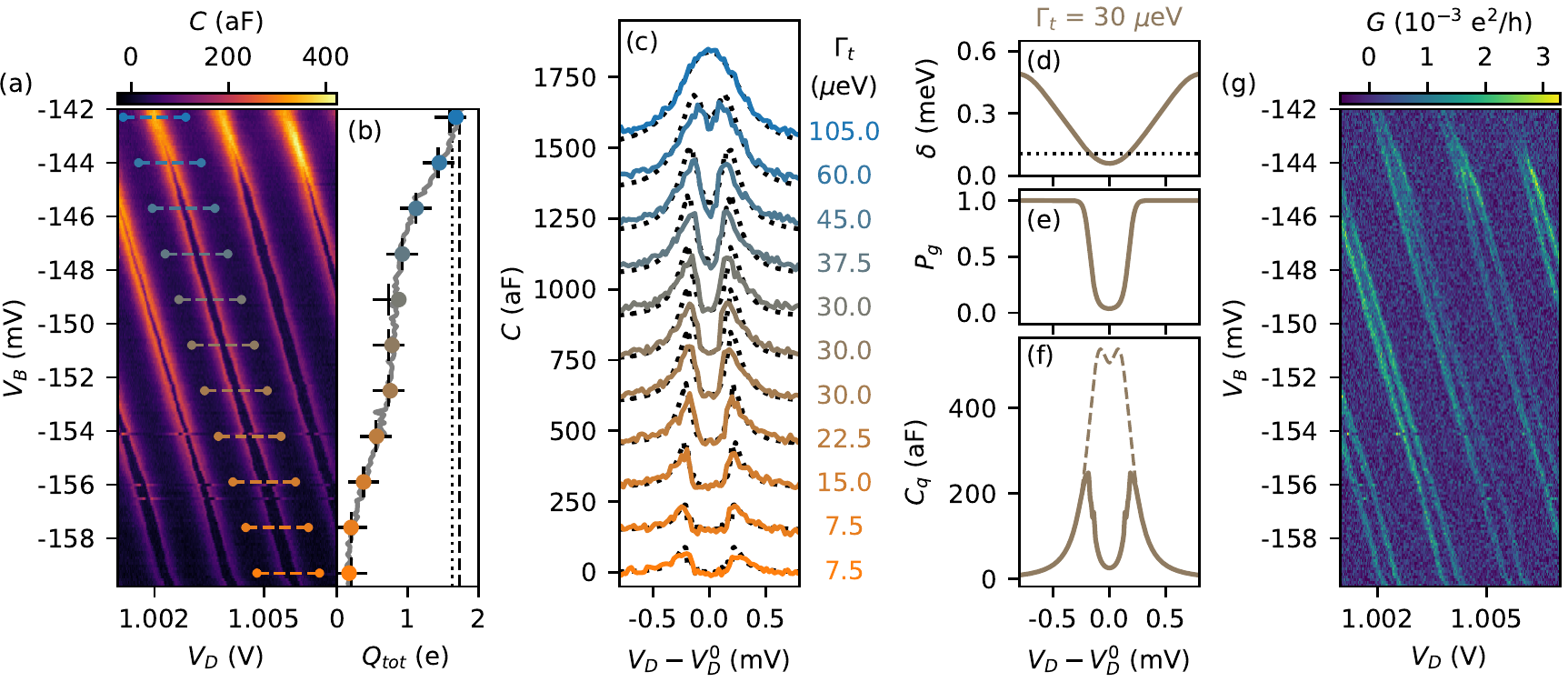}
	\caption{(a) Capacitance $C$ as a function of the barrier gate voltage $\Vb$, revealing the shrinking and vanishing of the charge stability regions. Dashed lines indicate the position of the cuts through the data (b,c), and the range of the integral $\Qt = \int \Cq d\Vd$.
	(b) Signature charge $\Qt$ of the charge transition (pair) as a function of $\Vb$. Colored points correspond to the line cuts and data in (a,c), and black crosses -- the corresponding signature charge extracted from the model. Dotted and dashed lines illustrate the value of $2 \alpha e$, with the lever arm $\alpha$ extracted from the measurement of the Coulomb diamonds (0.82), and the model fit (0.87), respectively.
	(c) Cuts through the data in (a). Solid, colored lines present the experimental data, and dotted black lines -- the fit result. Curves are offset vertically for clarity.
	(d,e) Excitation energy of the lowest excited state $\delta$ and ground state occupation probability $P_{g}$ extracted from the model and the fit for $\gamt=30$~\ueV.
	(f) Simulated quantum capacitance of the ground state for $\gamt=30$~\ueV (dashed), and the total quantum capacitance (solid) suppressed due to thermal excitation.
	(g) Conductance $G$ as a function of the barrier gate voltage $\Vb$, extracted from the same raw data as (a).
	}
	\label{fig_opening}
\end{figure*}

Fig.~\ref{fig_opening} presents a transition between the hybridization regimes in a single charge stability diagram
\footnote{Due to telegraph noise, presumably from nearby two-level system (TLS), Fig.~\ref{fig_opening}(b) is a composite of several repetitions of the measurement, all in an identical voltage range (see Appendix~\ref{app_compositing} for compositing procedure, Supplementary Fig.~\ref{fig_opening_2} for a complementary data for the other state of the TLS). Analogous data set in a regime with an additional subgap state in the SC island ($\Vi=0.25$~V) is presented in Supplementary Fig.~\ref{fig_opening_plunger}.}, measured with respect to dot and barrier gate voltages, $\Vd$ and $\Vb$. A range of $\Vb$ is chosen so that the barrier gate tunes the tunnel coupling with only relatively small change in $\Vb$. 
The shrinking and vanishing of the stability region \odd with increasing $\Vb$ is highlighted by taking line cuts through the charge stability diagram (Fig.~\ref{fig_opening}(c)). As the region \odd shrinks, the magnitude of capacitance $\C$ at the charge transition increases and becomes maximal when the pair of charge transitions merges.

We propose an intuitive understanding of the region \odd through an analogy to the singlet-doublet quantum phase
transition in the case of the QD coupled to a grounded SC~\cite{lee2014}. In that case, the QD charging energy $U$ competes with the tunnel coupling $\gamt$ to the SC lead. As the QD level $\varepsilon$ is tuned, the limit of $\gamt \ll U$ favors increments of the QD occupancy in steps of one electron, switching between the singlet states at even filling and doublet states at odd filling. In contrast -- large tunnel coupling ($\gamt \gg U$) favors increases of QD occupancy in steps of two electrons and the system remains in the singlet state at all times. In floating devices with fixed total charge the total system parity cannot change and therefore the quantum phase transition does not occur. Nonetheless, the parity of the QD may change provided a sufficient amount of thermal energy is available to excite a quasiparticle to the quasicontinuum of the SC states with high multiplicity $N$. As illustrated in Fig.~\ref{fig_device}(d), for sufficiently small tunnel coupling the quasicontinuum is separated from the discrete SGS by $\delta \! \lesssim \! \lnn k_B T$. 
Quasicontinuum states do not couple to the resonator, hence we interpret the sharp edges of the capacitance peaks, illustrated in Fig.~\ref{fig_opening}(a,c,f), to be due to such thermal excitations. The shrinking of region \odd is due to increasing $\delta$ for increasing $\gamt$. For sufficiently strong $\gamt$ relatively to charging energies, so that $\delta > \! \lnn kT$ for any value of $\Vd$, the charge transitions merge. In the following, we model the capacitance and its suppression to lend support to this interpretation.

We employ a model of a single-level Anderson impurity coupled to a finite-sized SC, with the Hamiltonian of the form~\cite{pavesic2021}
\begin{equation}
\begin{split}
	\hat{H} = \ & \varepsilon \hat{n} + U n_{\downarrow} n_{\uparrow} +
	\sum\limits_{k} \varepsilon_k \hat{n}_k + \frac{E_C^S}{2}\left(\sum\limits_{k} \hat{n}_k - n_g\right)^2\\
	- \ & \xi \sum\limits_{k,q} \left(  c_{k\uparrow}^\dag c_{k\downarrow}^\dag c_{q\downarrow} c_{q\uparrow}  + \mathrm{H.c.}\right)
	+ V \sum\limits_{k, \sigma} \left(  c_{k\sigma}^\dag d_{\sigma}  + \mathrm{H.c.}\right),
	\label{eq_H}
\end{split}
\end{equation}
where $d_{\sigma}$ and $\hat{n}_{(\sigma)}$ are annihilation and electron number operators for the impurity, $\sigma=\uparrow,\downarrow$. $c_{k\sigma}$ is the annihilation operator for island orbital $k$ with energy $\varepsilon_k$. $\varepsilon = \alpha \Vd$ is the energy of the impurity level; $\alpha$ -- the lever arm; $U$ -- the impurity charging energy; $E_C^S$ -- the SC island charging energy; $\xi$ -- the SC pairing strength; $V$ -- the impurity-bath hopping.
The model is solved using the DMRG method (Supplementary Sec.~\ref{app_dmrg}).
We compute the charge susceptibilities $\chi_{g,e}$ for the two lowest energy states (ground state $g$ and excited state $e$), separated in energy by $\delta$. We do not explicitly compute higher excited states, but instead assume high multiplicity $N$ for the excited state $e$~\cite{tinkham1995}. We expect $N \lesssim 10^8$, estimated based on the number of Al atoms composing the SC shell.

The quantum capacitance is given by
\begin{equation}
	\Cq = P_g \chi_g  + (1-P_g) \chi_e,
\end{equation}
where $P_g=(1+Ne^{-\delta / k_B T})^{-1}$ is the equilibrium occupancy of the ground state at temperature $T$. Since $\chi_e \ll \chi_g$ in the relevant gate voltage range (Supplementary Fig.~\ref{fig_opening_model}), for $\delta \lesssim \lnn k_B T$ the quantum capacitance is strongly suppressed.
Assuming the tunneling capacitance to be small, we fit the experimental data with a model that takes into account solely the quantum capacitance contribution, see Fig.~\ref{fig_opening}(c). In the simultaneous fit to all curves we use fixed values of $\Delta = 250$~\ueV, $U=502$~\ueV \ and $E_C^S = 196$~\ueV  (estimated from data presented in Suppl. Fig.~\ref{fig_diamonds}), common free parameters $\alpha$ and $T$, and separate free parameter $\gamt$ for each cut. Furthermore, the fit includes a constraint of $\gamt$ effectively enforcing its monotonous increase with $\Vb$. The fit results are presented with black dashed lines in Fig.~\ref{fig_opening}(c), and yield $\alpha=0.87$, $T=168$~mK and $N=1.6\times10^3$. Fig.~\ref{fig_opening}(d-f) illustrates for the case of $\gamt=30$~\ueV \ how a variable $\delta$ translates into $P_{g}$ that in turn determines the contribution of the ground-state quantum capacitance to the total value of $C$. 

In contrast, assuming no thermal excitations we are neither able to reproduce the magnitude of $\Cq$ for all data sets nor the degree of asymmetry for the most negative values of $\Vb$ (see Supplementary Fig.~\ref{fig_opening_model}). The obtained value of $T=168$~mK is larger than the base temperature of our setup ($30$~mK), which may be related to effectively increased temperature due to rf excitation, imprecision of describing the quantum dot as a single-level impurity, or strong covariance between $T$ and $N$ in the nonlinear fit. 

As a consistency check of our interpretation we calculate $\int \Cq d\Vd = \Qt$ across charge transition pairs, to yield a quantity we dub the ``charge signature'' of the transition (Fig.~\ref{fig_opening}(a,b); Supplementary Section~\ref{app_charge}).
In the absence of thermal screening we expect a charge signature $\Qt = 2 \alpha e$, reflecting the transfer of two electrons from the QD to the SC island. Indeed, with vanishing region \odd for most positive $\Vb$ the value of $\Qt$ approaches about $1.7e$ consistent with $\alpha$ extracted from the fit in Fig.~\ref{fig_opening}(b) and from the Coulomb diamond measurement of the dot (dashed and dotted line, respectively; Supplementary Fig.~\ref{fig_diamonds}). Conversely, for more negative $\Vb$, as the region \odd increases in size $\Qt$ decreases significantly, in agreement with the interpretation of the $\Cq$ suppression. We note that in other data sets in which we observe merging of the charge transitions the charge signature reaches different maximal values, usually smaller. We suspect this is related to different lever arms in the different configurations, but currently lack supporting evidence.


The proposed model is sufficient for modeling capacitance, however it neglects the tunneling capacitance. We conjecture that the tunneling capacitance is much smaller than quantum capacitance, but associated relaxation processes lead to measurable Sisyphus resistance~\cite{ashoori1992,esterli2019}, which is revealed as enhanced dissipation in the resonator. Fig.~\ref{fig_opening}(g) presents the dissipation at the charge transitions, expressed as the conductance $G$ extracted from the same data as the capacitance $C$ in Fig.~\ref{fig_opening}(a,c). For the most negative $\Vb$, the peaks in $G$ coincide with those in $C$. Unlike the $C$ peaks however, the $G$ peaks do not quite merge for the most positive $\Vb$. Supplementary Figs.~\ref{fig_opening_2} and \ref{fig_more_maps} further indicate that when the $\Vb$ is increased the peaks vanish rather than merge. The double peak structure resembles the result of Ref.~\cite{esterli2019} that predicts a double-peak for the Sisyphus resistance at the interdot charge transition for a single-electron double QD. As a partial validation of the hypothesis, we extend the model by considering a relaxation rate $\Gamma$ and calculate the dissipation related to transitions between ground and excited states over one period of the rf drive. Supplementary Fig.~\ref{fig_opening_model} demonstrates that we are able to reproduce $G$ of the correct order of magnitude and that it furthermore exhibits a double-peak structure. We refrain from further comparison to the experimental data, since the relaxation rate $\Gamma$ is strongly dependent on both tunnel coupling and dot-island detuning, which we did not study experimentally.


To conclude we outline possible directions of research building on our study of a subgap state in a QD coupled to a SC island by means of dispersive gate sensing. Since gate sensing has already enabled the development of large dot arrays and multiqubit manipulation, we envision that dispersive gate sensing will likewise enable the tuning of larger arrays and chains of hybrid SC devices, which so far were two-~\cite{deacon2015,su2017,saldana2018,saldana2020,vekris2022,dvir2022} or at most three-sites long~\cite{wu2021}. A particularly appealing direction to pursue is the construction of synthetic Kitaev chains with topological ground states~\cite{kitaev2001,sau2012,fulga2013,dvir2022}. By controlling the potential along the chain with the help of gate sensing one could measure the site-resolved local compressibility, which is an equivalent of the quantum capacitance, and is predicted to peak at the topological phase transition~\cite{nozadze2016}.

\section*{Author contributions}

F.K.M. performed the experiment with input from L.H., D.J., M.C., C.P. and J.V.K.
F.K.M. processed the experimental data.
L.P. and R.\v{Z}. solved the Anderson impurity model in DMRG framework.
R.K.R., Z.G., C.P. and  A.P. developed framework describing the thermal screening and dissipation, and performed the fits to the data.
Y.L. and P.K. grew the InAs nanowires with an epitaxial aluminum shell.
D.J. fabricated the nanowire device.
R.\v{Z}. and A.P. supervised the theory.
F.K.M. wrote the manuscript with input from R.\v{Z}., J.V.K., R.K.R., Z.G., L.P., C.P., L.H., A.P., M.C.

\section*{Acknowledgments}

This work is supported by the Netherlands Organization for Scientific Research (NWO),
Microsoft Quantum Lab Delft, Quantum Materials Lab Copenhagen, and Quantum Lab Sydney.
L.P. and R.\v{Z}. acknowledge the support of the Slovenian Research Agency (ARRS) under Grants No. P1-0044, P1-0416, and J1-3008.
Z.G. and A.P. were supported by the Hungarian Ministry of Innovation and Technology and the National Research, Development and Innovation Office (NKFIH) within the Quantum Information National Laboratory of Hungary, and by the NKFIH through the OTKA Grant FK 132146.
F.K.M. acknowledges support from NWO under a Veni grant (VI.Veni.202.034).

%

\appendix
\counterwithin{figure}{section}

\section{Resonator model and extraction of $C$ and $G$}
\label{app_resonator}

To extract the QD contributions to the resonator impedance, specifically the capacitance $C$ and the conductance $\G$, we measure the reflection coefficient of a resonant circuit comprised of a SC spiral inductor ($L = 570$~nH), parasitic capacitance to ground, and the dot-island device itself. As illustrated in Fig.~\ref{fig_device}(b) the resonator losses and parasitic capacitance are treated as being in parallel to $\C$ and $\G$ of the device. The extraction of $\C$ and $\G$ requires the following:

\begin{enumerate}
	\item an analytical resonator model, describing the complex reflection coefficient $\S$ as a function of the resonance frequency $\fres$, internal quality factor $\Qi$ and number of other parameters;
	\item an analysis procedure for extracting $\fres$ and $\Qi$ from the measurement of $\S$:
	\begin{itemize}
		\item if the dependence of $\S(f)$ is measured explicitly, this role is performed by a nonlinear fit to the data;
		\item in case of the fixed-frequency measurement at known probing frequency $f$, it is achieved using the mapping $\S \mapsto \fres, \Qi$, generated numerically based on the resonator model and an individual measurement of $\S(f)$;
	\end{itemize}	 
	\item conversion of changes in $\fres$ and $\Qi$ to $\C$ and $\G$.
\end{enumerate}

\subsection{Resonator model}

In our work we employ the resonator model described in detail in Ref.~\cite{malinowski2022}, and we only summarize it here. The model consists of three parts.

First, the resonator itself, modeled as a RLC resonator (with $L$ and $C$ arranged in series) coupled to a 50~$\Omega$ transmission line. Its reflection coefficient $\S'$ is given by
\begin{equation}
	\label{eq_RLC}
	\S'(f) = 1 - \frac{ 2 \Ql \Qe^{-1} }{1 + 2 i \Ql \frac{(f - \fres) }{\fres}},
\end{equation}
where $f$ is the probing frequency, $\Ql=(\Qi^{-1} + \Qe^{-1})^{-1}$ is the loaded quality factor, $\Qi$ is the internal quality factor, $\Qe$ is the external quality factor, and $\fres$ is the resonance frequency.

The second element is the low-quality cavity formed between the resonator and the partially reflective input of the cryogenic amplifier. The cavity modifies the measured signal, and can lead to asymmetry in the resonance. The modified reflection coefficient is given by

\begin{equation}
	\label{eq_resonator}
	\S'' = \frac{i \gamma \alpha \sqrt{1-\gamma^2} e^{-4 \pi f l i / c + i\phi} \S'}{1 - \sqrt{1-\alpha^2} (1-\gamma^2) e^{-4 \pi f l i / c + i\phi} \S'},
\end{equation}
where $\gamma$ is the coupling coefficient of the used directional coupler, $\alpha$ is amplifier reflection coefficient, $l$ is length of the coaxial cables connecting the resonator with the amplifier, and $c$ is propagation speed of the rf excitation in the coaxial cables.

Finally, global phase winding, phase offset and frequency-dependent amplification are taken into account, modifying the measured reflection coefficient~\cite{probst2015}
\begin{equation}
	\label{eq_khalil_extended}
	\S = A \left( 1+B \frac{f - \fres}{\fres} \right) \times e^{-i a+i b(f-\fres)} \times \S'',
\end{equation}
where $A \left( 1+B \frac{f - \fres}{\fres} \right)$ accounts for a frequency-dependent attenuation and amplification, while $e^{-i a+ i b(f-\fres)}$ accounts for the phase winding and offset.

For the procedure of fixing the numerous parameters in the model we refer the reader to the appendix of Ref.~\cite{malinowski2022}.

\subsection{Conversion of fixed-frequency measurement to $\fres$ and $\Qi$}

\begin{figure*}[tbh]
	\includegraphics[scale=1]{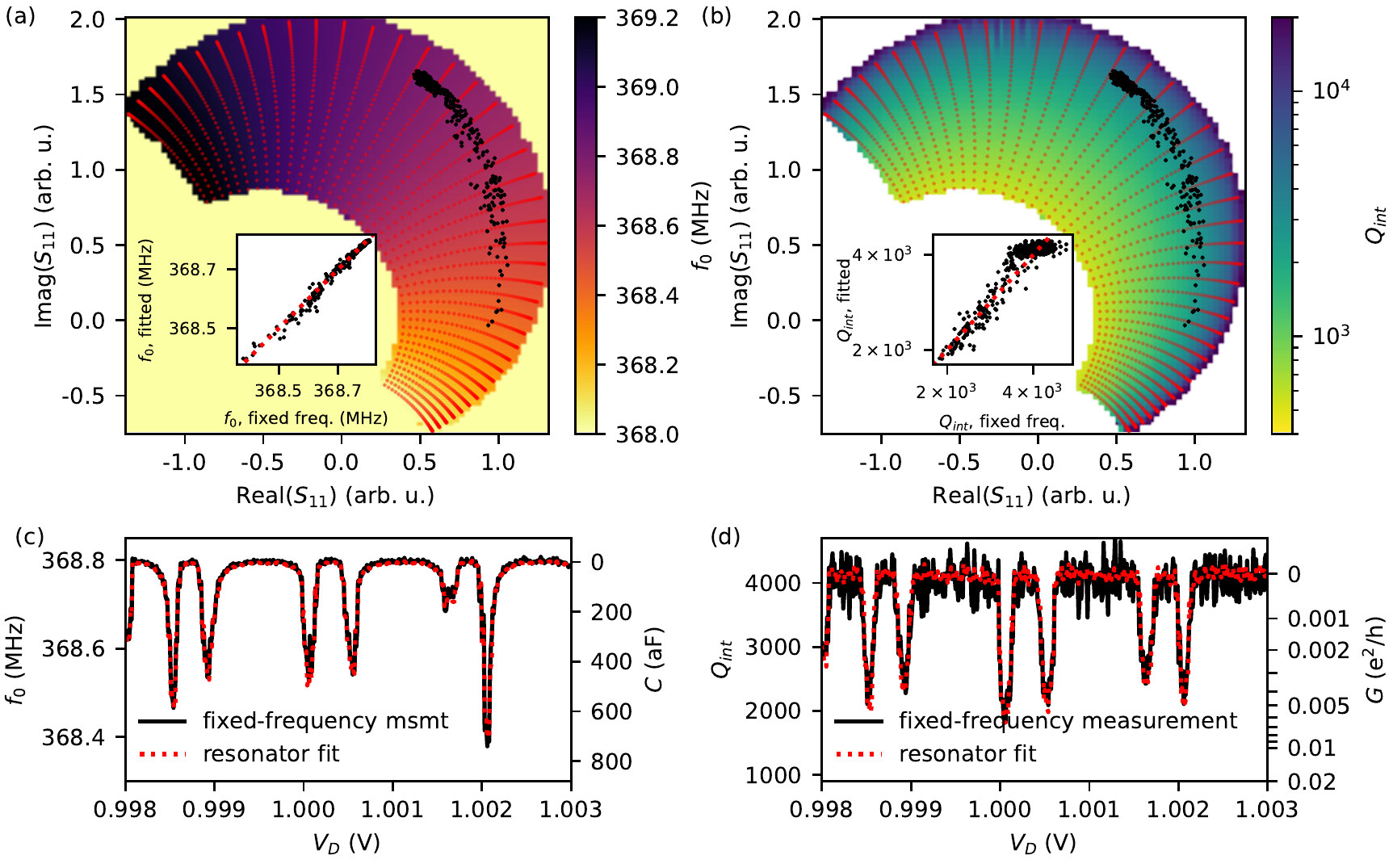}
	\caption{(a,b) Mapping of the fixed-frequency measurement of $\S$ to $\fres$ and $\Qi$, respectively. Red points indicate the grid of ($\fres$, $\Qi$) used to generate the mapping. The actual grid used has 8 times more points than shown. Black points indicate the scattered fixed-frequency data from the data set corresponding to the green curve in Fig.~\ref{fig_subgap}(d)
	(c,d) Overlayed values of $\fres$ and $\Qi$, extracted from the full resonator fit and the fixed-frequency measurements.
	Insets of (a,b) show correlations between the data sets in (c,d), with red dashed lines indicating $x=y$.}
	\label{fig_mapping}
\end{figure*}

While the majority of the presented data is derived from the full resonator fit to the frequency-dependent $\S$, the color maps shown rely on fixed-frequency measurement at the probing frequency $f=368.4$~MHz. In order to convert an individual measurement of the complex value of $\S$ to $\fres$ and $\Qi$ we apply the following procedure:

\begin{enumerate}
	\item Perform an individual, high-resolution measurement of $\S(f)$ and fit it with an analytical resonator model.
	\item Fix all of the model parameters except for $\fres$ and $\Qi$.
	\item Evaluate the expected value of $\S$ for a dense grid of ($\fres$, $\Qi$) values.
	\item Numerically invert the $\S(\fres, \Qi)$ function to generate the mapping $\S \mapsto (\fres, \Qi)$.
	\item Verify the uniqueness of the mapping in the relevant range of $\fres$ and $\Qi$ values.
	\item Apply the mapping to the measured $\S$ at a fixed-frequency $f$.
\end{enumerate}

Fig.~\ref{fig_mapping} illustrates an example of the mapping $\S \mapsto (\fres, \Qi)$, based on a single resonator measurement from the full data set, corresponding to the undepleted island as in Fig.~\ref{fig_subgap}(d). In Fig.~\ref{fig_mapping}(a,b), color maps illustrate the mapping $\S \mapsto (\fres, \Qi)$, red points illustrate the grid of the $(\fres, \Qi)$ points, and black points are a scatter plot of $\S$ values measured at fixed probing frequency $f$, while tuning $\Vd$. We note that due to the resonator asymmetry there is no range of $\fres$, $\Qi$ for which the mapping $\S \mapsto (\fres, \Qi)$ can be separated into two independent mappings, $|\S| \mapsto \Qi$ and $\angle(\S) \mapsto \fres$, and thereby the signal magnitude and phase do not independently represent resistance and capacitance.

In figure \ref{fig_mapping}(c,d) we compare the values of $\fres$ and $\Qi$ extracted from the full resonator fit and from the mapping of individual values of $\S(f)$ at $f=368.4$~MHz, from the same data set. We find agreement between the two methods, albeit with much higher noise in the case of fixed-frequency measurements. The correlation between the $\fres$ and $\Qi$ values extracted with the two methods are presented in the insets of Fig.~\ref{fig_mapping}(a,b). An excellent correlation justifies the use of this extraction method of $\fres$ and $\Qi$ from the fixed-frequency measurement of $\S$ [(Fig.~\ref{fig_subgap}(a-c) and \ref{fig_subgap_G}(a-c)].

\subsection{Conversion of $\fres$ and $\Qi$ to $\C$ and $\G$}

Having converted the measured reflection coefficients to $\fres$ and $\Qi$, we establish their reference values, i.e. the values $\fref$ and $\Qi^{ref}$ corresponding to $\C=0$ and $\G=0$. For this purpose we use the mean of the values deep in the Coulomb blockade. Subsequently we calculate
\begin{equation}
	C_q = \frac{1}{(2 \pi \fref)^2 L} - \frac{1}{(2 \pi \fres)^2 L},
\end{equation}
using the nominal value of $L=570$~nH.

We note that the value of $C_q$ extracted in this way is susceptible to a significant source of potential systematic error. Namely, if the width of the charge transitions becomes comparable to the spacing between them, the quantum capacitance in the middle of the stability region may be nonzero. This can lead to $C_q$ values overestimated by a constant up to about $100$~aF, different for each data set. This has a particularly important implication for the calculation of the charge signature $\Qt$, where an incorrect calibration of $\fref$ can compound to a gross underestimation. In the data we present to support the claim that $\Qt$ saturates at $2 \alpha e$ (Fig.~\ref{fig_opening}), the value of $\fref$ is extracted from the 2D data set in Fig.~\ref{fig_opening}(a), and in particular -- based on the most negative values of $\Vb$. There, we expect the $\fref$ can be extracted most reliably. Relative to that background, the value of $C_q$ increases up to 60-70~aF. If this offset is miscalibrated, the value of $\Qt$ may be underestimated by as much as bout $70 \mathrm{aF} \times 1.7 \mathrm{mV} \approx 0.7 e$.

To obtain $\G$ we first calculate the characteristic impedance of the resonator $Z_{ch} = 2 \pi L \fref$. For a resonator model as in Fig.~\ref{fig_device}(b) the total conductance of the two resistors to ground (one representing losses in the resonator, another in the dot-island system) is given by $G^{tot} = 1/ Z_{ch} \Qi$. Similar to the case of capacitance, we identify the value of conductance in Coulomb blockade, $G^{ref}$, as representing the losses intrinsic to the resonator. Any increase of conductance above the reference level we attribute to the contribution of dissipation in the dot-island: $\G = G^{tot} - G^{ref}$.

\section{DMRG calculation}
\label{app_dmrg}

The DMRG used to solve the Anderson impurity model of a QD coupled to a SC island is described in Ref.~\cite{pavesic2021}.
The number of island orbitals $\mathcal{L}$ is set to 200; further increase in number does not produce significant gains in terms of accuracy. All parameter values are given in the units of half-bandwidth $D=1$. For all terms in the Hamiltonian to have the same operator norm, certain interactions have to be rescaled by the system size. The one-to-all impurity hopping is thus $V = v/\sqrt{\mathcal{L}}$. Hopping strength is further referred to in terms of $\mathcal{L}$-independent tunneling rate $\gamt = \pi \rho v^2$, where $\rho$ is the normal-state density of states.
The all-to-all SC pairing strength is $\xi = x/\mathcal{L}$, where we choose $x=0.4$. This determines the SC gap $\Delta = 0.165$, which allows us to relate the calculation to experiment. $x$ is chosen such that an appropriate number of levels participate in SC pairing while still minimizing the effect of finite bandwidth. 

The Hamiltonian \eqref{eq_H} can be represented in the MPO (matrix product operator) form with $9 \times 9$ matrices, which allows for efficient and exact calculations. 
The maximal matrix dimension of the matrix product state during optimization sweeps is 3000, while the energy cutoff (the lowest Schmidt value retained during the singular value decomposition step) is $10^{-10}$. Such calculations take on the order of a few hours and ensure very good convergence of all physical properties. 

The output of the DMRG calculation are the energies of the ground and first excited state and their QD occupations. The occupations are used to obtain the capacitance as detailed in Appendix \ref{app_capacitance}, while the energy difference between the two states enters the Boltzmann weight in thermal screening. 

The code with examples is available on Zenodo~\cite{QD_SI_solver}. 

\section{Derivation of the effective capacitance and resistance from DMRG results}
\label{app_capacitance}

\begin{figure*}[tbh]
	\includegraphics[scale=1]{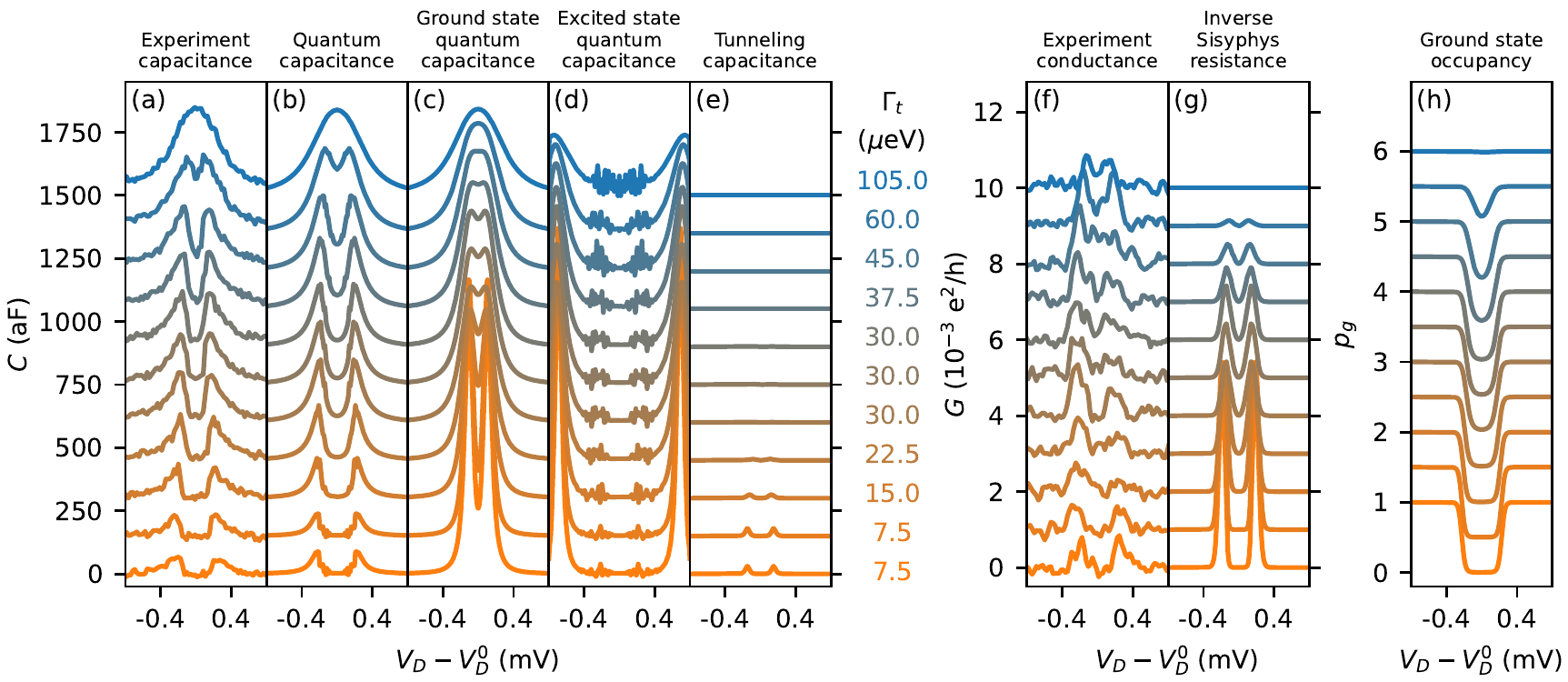}
	\caption{Reproduction of the data from Fig.~\ref{fig_opening}, alongside the intermediate output of the model.
	(a) Measured capacitance;
	(b) Total quantum capacitance fitted to the data (identical to dashed lines in Fig.~\ref{fig_opening}(c));
	(c) Quantum capacitance of the ground state;
	(d) Quantum capacitance of the lowest excited states. The noise in the data (visible also in (b)) corresponds to numerical error, due to large number of nearly-degenerate states posing a challenge to DMRG optimization.
	(e) Tunneling capacitance extracted from the model using zero-temperature relaxation rate of $\Gamma = 32 \times 10^6$~s$^{-1}$;
	(f) Cuts through the color map of $\G$ in Fig.~\ref{fig_opening}(g), in the same range as the cuts (a). To suppress the noise, the data in the color map was convolved with a 2D Gaussian kernel, strongly elongated in along the charge transition features.
	(g) Inverse Sisyphus resistance extracted from the model using $\Gamma = 32 \times 10^6$~s$^{-1}$;
	(h) Thermal probability of the ground state occupancy.
	}
	\label{fig_opening_model}
\end{figure*}

In this section we derive the response of the charge accumulated on the dot plunger gate as a result of the dot charge responding to small cosine voltage excitation, and interpret it as an effective capacitance and resistance loading the radio frequency resonator, analogously to Ref.~\cite{esterli2019}.

First, we assume that the excited states can be effectively described by a collection of $N$ degenerate states, at the energy of the lowest excited state. Fig.~\ref{fig_opening_model}, summarizing the outputs of DMRG calculation and the calculations below, illustrates that in the regime where the population of excited states is appreciable the quantum capacitance of the lowest excited state is small, and the higher excited states are only expected to have lower capacitance. This is the critical observation leading to a thermal suppression of capacitance, and justifying the simplified treatment of excited state as $N$-fold degenerate.

Under this assumption, charge accumulated on the gate is
\begin{equation}
\label{eq:Qt}
    Q = \alpha |e| (P_g n_g + P_e n_e)
\end{equation}
where $P_{g/e}$ is the occupation probability of the ground and excited state, $n_{g/e}$ is the dot charge in the ground and excited state, $\alpha$ is the plunger-gate lever arm and $e$ is the unit charge (negative). The functional form of $n_g$ and $n_e$ versus plunger gate voltage is obtained by the DMRG calculation described in the previous section (Sec.~\ref{app_dmrg}).

We assume that the plunger gate voltage has a DC and an AC part $\Vd(t) = \Vd^0+\Vd^{AC} \sin(\omega t)$, where $\Vd^0$ is a DC offset while $\Vd^{AC}$ is the amplitude of the driving AC signal. Next, we obtain $P_g(t)$ by solving the master equation governing the time evolution of the occupation probabilities
\begin{equation}
	\label{eq_diffeq}
	\dot{P}_g(t) = - N \Gamma_{\uparrow}(t) P_g(t) 
	+ \Gamma_{\downarrow}(1-P_g(t)),
\end{equation}
where $\Gamma_{\uparrow}(t)$ ($\Gamma_{\downarrow}(t)$) is the uphill (downhill) relaxation rate
\begin{equation}
	\begin{aligned}
		\Gamma_\downarrow(t) &\equiv
		\Gamma_\downarrow(\delta(t)) = 
		\Gamma \left(1+ n_{B}(\delta(t),T) \right), \\
		\Gamma_\uparrow(t) &\equiv
		\Gamma_\uparrow(\delta(t)) = 
		\Gamma n_{B}(\delta(t),T), 
	\end{aligned}
\end{equation}
$n_B$ is the Bose-Einstein function
\begin{equation}
	n_B(\delta,T) = \frac{1}{e^{\delta/k_B T} -1},
\end{equation}
$\Gamma$ is the zero-temperature downhill relaxation rate, and $\delta(\Vd(t))$ is the splitting between the ground and the excited state.

Expanding $\Gamma_{\downarrow,\uparrow}$, $n_{g,e}$ and $\delta$ linearly around $\Vd^0$ we look for linear-response steady-state solutions of the differential equation (Eq.~\ref{eq_diffeq}) of the form
\begin{equation}
	\label{eq_parallelRCcharge}
	Q(t) = C \sin(\omega t) V_D^{AC} - \frac{1}{R\omega} \cos(\omega t) V_D^{AC}.
\end{equation}

In the solution we identify the resistance and capacitance which may be expressed as
\begin{equation}
	\label{eq_C}
	\begin{aligned}
		C &=- \alpha^2 P_{g0} |e|^2 n_g'  -\alpha^2 (1- P_{g0})|e|^2 n_e' \\
		&+
		\alpha^2(n_{e0}-n_{g0}) |e|^2 
		\frac{- \left[ P_{g0} N \Gamma_{\uparrow}'- (  1-P_{g0} ) \Gamma_{\downarrow}' \right] }{\omega^2+ \left( N \Gamma_{\uparrow 0} + \Gamma_{\downarrow 0} \right)^2} \left( N \Gamma_{\uparrow 0} + \Gamma_{\downarrow 0} \right),
	\end{aligned}
\end{equation}
\begin{equation}
	\label{eq_R}
	\frac{1}{R} = \alpha^2 (n_{e0}-n_{g0}) |e|^2 \frac{- \left[ P_{g0} N \Gamma_{\uparrow}'- (  1-P_{g0} ) \Gamma_{\downarrow}' \right]}{\omega^2+ \left( N \Gamma_{\uparrow 0} + \Gamma_{\downarrow 0} \right)^2}
\omega^2.
\end{equation}
where symbol $'$ denotes derivative ${d \bullet}/{d \epsilon} \at[\Big]{\epsilon_0}$ and subscript ``$0$'' denotes value of the parameter at $\epsilon = \epsilon_0 = \alpha \Vd^0$.

In Eq.~\ref{eq_C} we identify the first line as the quantum capacitance, and the second one as the tunneling capacitance. Only the tunneling capacitance depends on the relaxation rates that are not explicitly known (they likely depend in a complex manner on $\epsilon$, $\gamt$ and $\delta$). Therefore, as mentioned in the main text, we perform the fit to the data only using the quantum capacitance contribution, and infer the presence of tunneling capacitance indirectly, through qualitative features, similar between the measured effective conductance and inverse of the calculated effective resistance (Eq.~\ref{eq_R}). This is a posteriori justified by a successful fit and the observation that the tunneling capacitance may indeed be assumed smaller than the quantum capacitance contribution.

\section{Charge signature of the transitions}
\label{app_charge}

In Fig.~\ref{fig_opening}(c) we integrate numerically the measured capacitance from the center of the charge transition (pair) $\Vd^0$ to $\Vd$
\begin{equation}
	Q(\Vd) = \int\limits_{\Vd^0}^{\Vd} \C(\tilde{V}_D) \mathrm{d} \tilde{V}_D,
\end{equation}
and in Fig.~\ref{fig_opening}(d) we present the integral of capacitance across the full charge transition pair $\Qt$. This quantities will be used to quantify to what extent the system is trapped in a state that does not contribute to the measured capacitance.

In the effective capacitance picture, $\C$ consists of two contributions -- so-called quantum and tunneling capacitance~\cite{esterli2019}.

Quantum capacitance is the contribution that arises from the adiabatic response of the system to the oscillating AC voltage on the gate attached to the resonator. One can consider this contribution, by considering each eigenstate of the system, labeled by subscript $i$, with energy $E_i(\Vd)$, to be occupied with certain probability $p_i(\Vd)$. Collectively they contribute
\begin{equation}
	\Cq(\Vd) = \sum\limits_{i} p_i \frac{\mathrm{d}^2 E}{\mathrm{d} \Vd^2} = \sum\limits_{i} p_i \frac{\mathrm{d} Q_i}{\mathrm{d} \Vd} = \alpha \sum\limits_{i} p_i(\Vd) C_i.
\end{equation}
	\label{eq_cq_general}
Here, $Q_i(\Vd)$ represents the expected value of charge on a dot for $i$-th eigenstate, $\alpha$ -- a lever arm between the dot and the gate, and $C_i$ -- capacitance associated with each eigenstate. In particular, if the ground state is occupied with probability $p_0(\Vd)=1$ (at zero temperature), it immediately follows that
\begin{equation}
	\int\limits_{\Vd^A}^{{\Vd^B}} C_q(\tilde{V)_D} \mathrm{d}\tilde{V}_D = \alpha(Q_0(\Vd^B)-Q_0(\Vd^A)).
\end{equation}
In particular, integrating between the middle of the stability regions with charge different by $2e$ should yield $\Qt=2 \alpha e$.

On the other hand, the tunneling capacitance results from redistribution of $p_i$ between the eigenstates over a single period of an AC excitation, and is affected by a rate at which the system reaches the thermal equilibrium, relative to the drive frequency. In particular, in the limit of the equilibration rate being much greater than the drive frequency
\begin{equation}
	C(\Vd) = \alpha \frac{\mathrm{d}\langle Q(\Vd) \rangle}{\mathrm{d} \Vd},
\end{equation}
where $\langle Q(\Vd) \rangle$ represents the average dot charge in thermal equilibrium. Also in this case integrating between the middle of the stability regions with charge different by $2e$ should yield $\Qt=2 \alpha e$.

For our experiment, we conclude that the reduction of $\Qt$ below value of $2 \alpha e$ can be attributed to the system being trapped in an excited state, and unable to respond to the gate voltage changes at a timescale compared to the period of a drive frequency. Gradual increase of $\Qt$ from nearly~0 to $\Qt \approx 2 \alpha e$ shows that the suppression of the signal by excitation of the system to quasi-continuum occurs to the lesser extent. As illustrated in Fig.~\ref{fig_device}(d), this is due to the tunnel coupling increasing the energy gap between the discrete and the quasi-continuum, making the limit $T \rightarrow 0$ increasingly adequate.

\begin{figure*}[tbh]
	\includegraphics[scale=1]{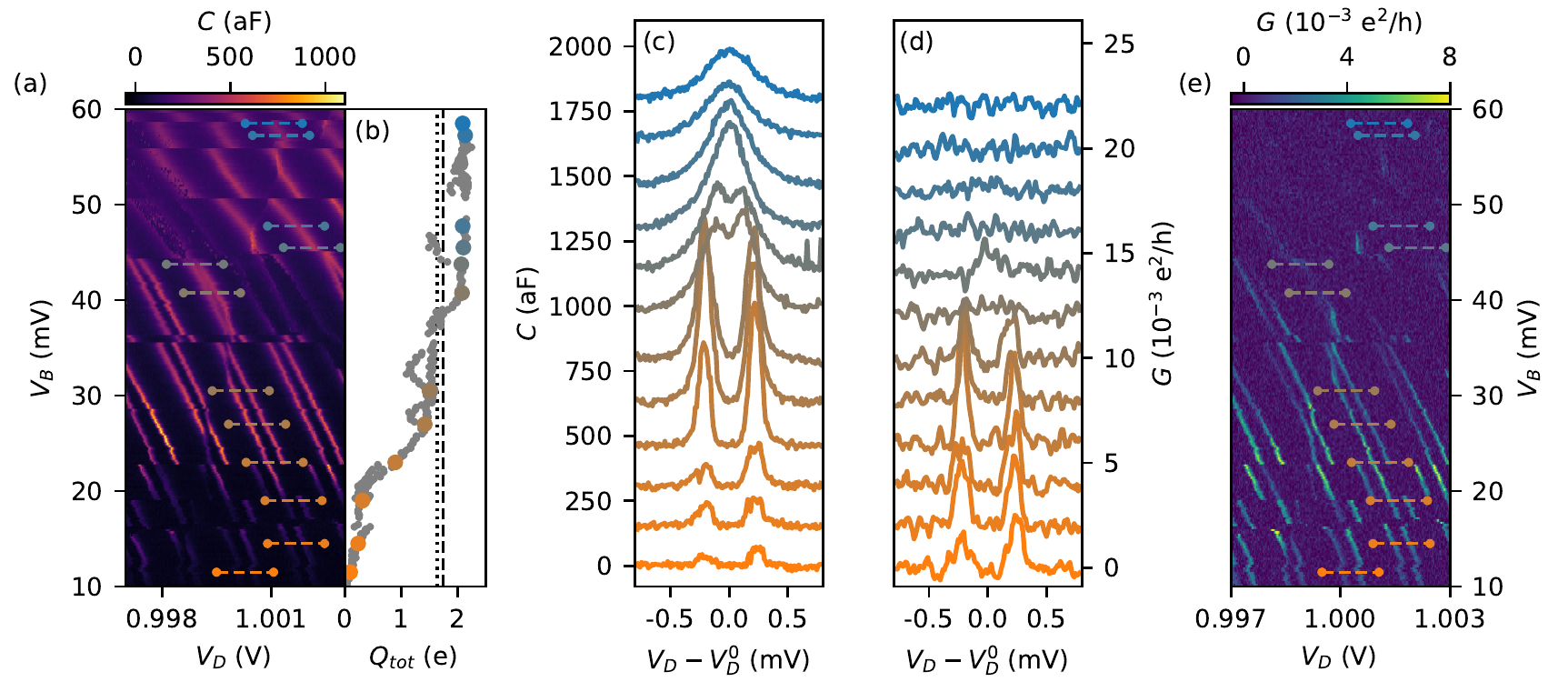}
	\caption{
	A data analogous to Fig.~\ref{fig_opening}, in for the island plunger gate voltage set to $\Vi=0.25$~V to accumulate electrons away from the Al shell, and create an additional subgap state.
	(a) A capacitance of the floating dot-island as a function of the barrier gate voltage $\Vb$. Dashed lines indicate position of the cuts through the data (b-d).
	(b) Signature charge $\Qt$ of the charge transition (pair) as a function of $\Vb$. Colored points correspond to the line cuts and data in (a,c-e) Dotted and dashed line line illustrates the value of $2 \alpha e$, with the lever arm $\alpha$ extracted from the measurement of the Coulomb diamonds, and model fit, respectively.
	(c) Cuts through the data in (a).
	(d) Cuts through the effective conductance data in (c.f. (e)). Before taking the cuts we apply a smoothing Gaussian filter, elongated in the direction along the charge transitions in order to increase a SNR.
	(e) An effective conductance of the floating dot-island as a function of the barrier gate voltage $\Vb$, extracted from the same raw data as (a). Dashed lines indicate position of the cuts in (b-d).
	}
	\label{fig_opening_plunger}
\end{figure*}



\section{Compositing procedure for barrier gate sweep}
\label{app_compositing}

\begin{figure*}[tbh]
	\includegraphics[scale=1]{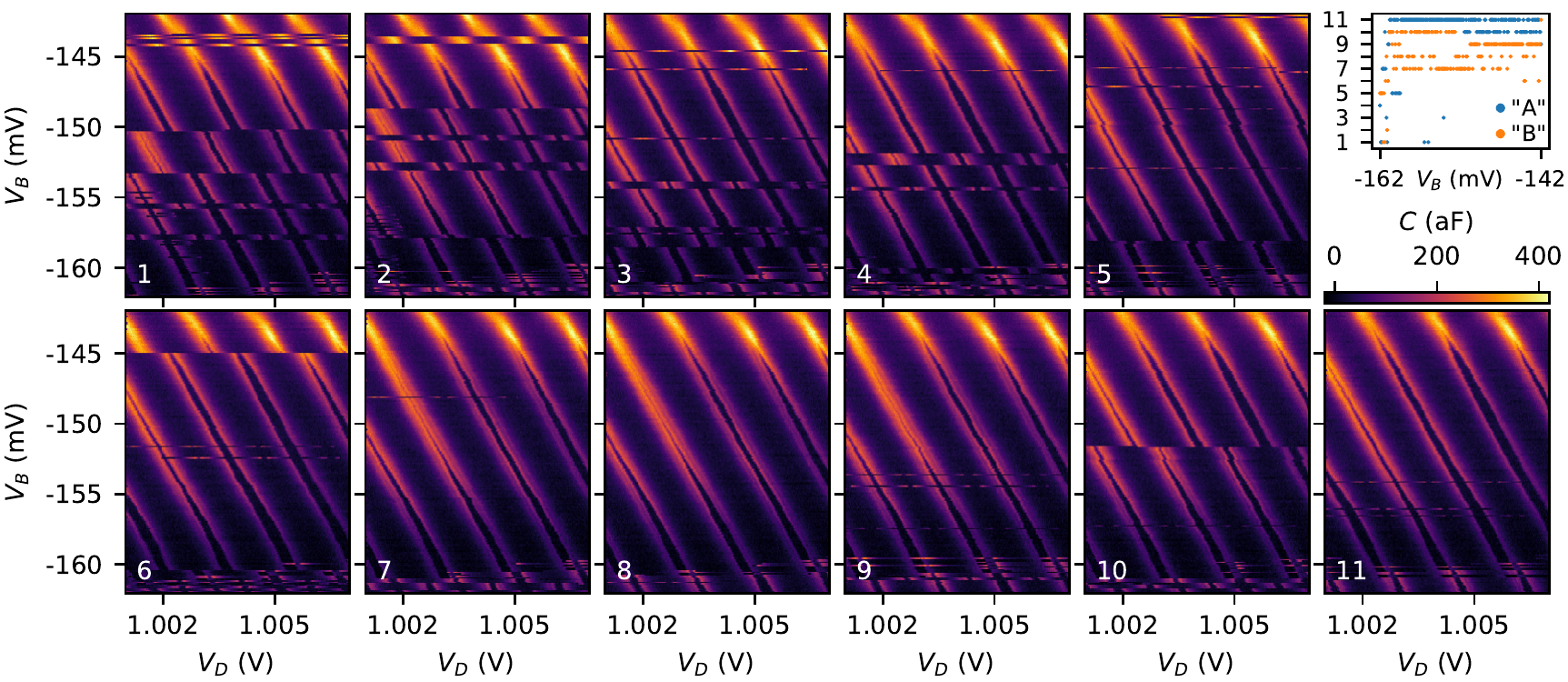}
	\caption{Summary of the capacitance measurement from the data sets that were used to create a composite data in Fig.~\ref{fig_opening} and Fig.~\ref{fig_opening_2}. Top tight panel summarizes which lines were included in Fig.~\ref{fig_opening} (``A'') and Fig.~\ref{fig_opening_2} (``B'').
	}
	\label{fig_composite}
\end{figure*}

\begin{figure*}[tbh]
	\includegraphics[scale=1]{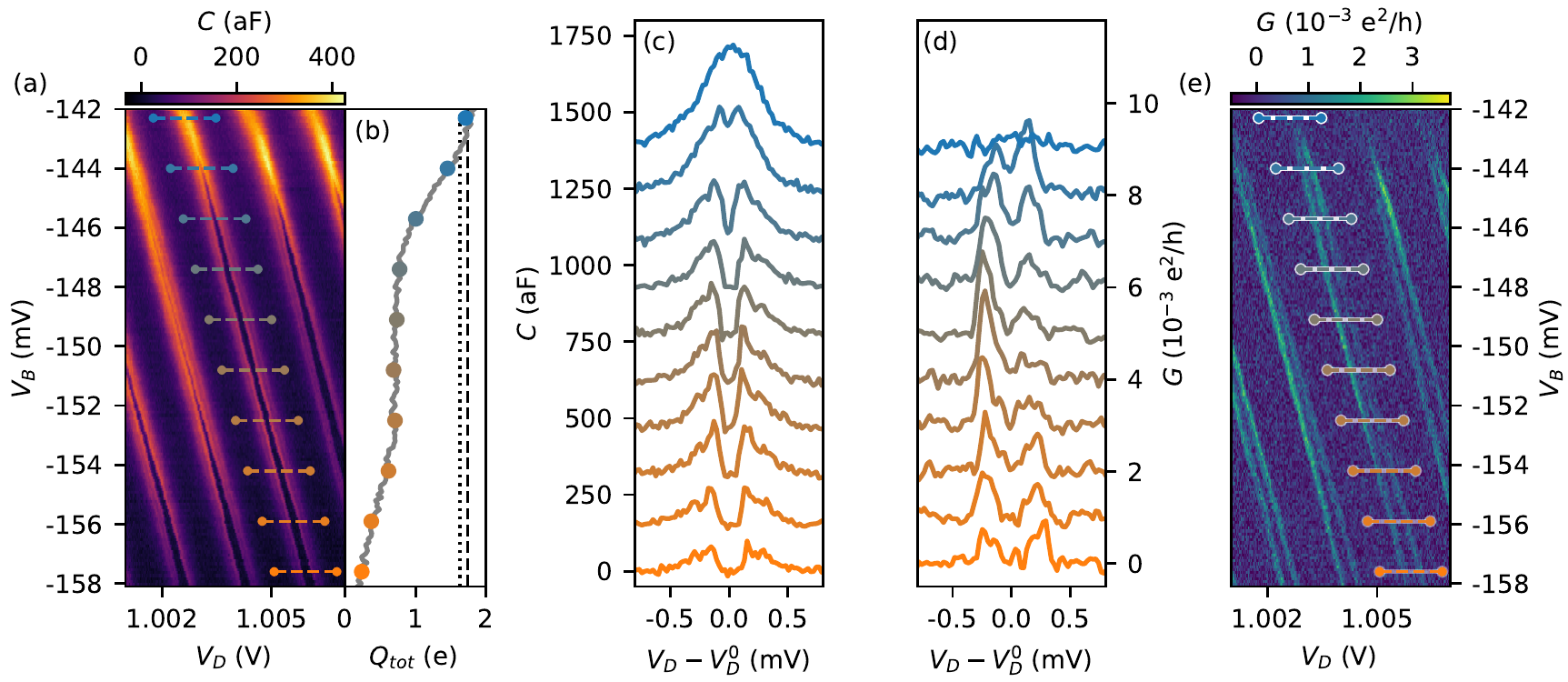}
	\caption{A data complementary to Fig.~\ref{fig_opening}, for the second state of the TLS.
	(a) A capacitance of the floating dot-island as a function of the barrier gate voltage $\Vb$, illustrating shrinking and vanishing of the charge stability regions. Dashed lines indicate position of the cuts through the data (b-d).
	(b) Signature charge $\Qt$ of the charge transition (pair) as a function of $\Vb$. Colored points correspond to the line cuts and data in (a,c-e) Dotted and dashed line line illustrates the value of $2 \alpha e$, with the lever arm $\alpha$ extracted from the measurement of the Coulomb diamonds, and model fit, respectively.
	(c) Cuts through the data in (a).
	(d) Cuts through the effective conductance data in (c.f. (e)). Before taking the cuts we apply a smoothing Gaussian filter, elongated in the direction along the charge transitions in order to increase a SNR.
	(e) An effective conductance of the floating dot-island as a function of the barrier gate voltage $\Vb$, extracted from the same raw data as (a). Dashed lines indicate position of the cuts in (b-d).
	}
	\label{fig_opening_2}
\end{figure*}

Due to poor device stability it was not possible to acquire the data set equivalent to Fig.~\ref{fig_opening} without switches line-to-line. To overcome it we identified a bistable region, where we judged the tunnel coupling to change nearly monotonously. In that range we performed 11 identical sweeps (Fig.~\ref{fig_composite}) that were then combined to generate two data sets with switches eliminated -- Fig.~\ref{fig_opening} and Fig.~\ref{fig_opening_2}.

The compositing was performed line-by-line, along the $\Vb$ axis. We start with the top line from an arbitrarily chosen data set. Then as $(i+1)$st line we select the line from the data set that has the maximum correlation with the previous, $i$ th, line.

The composite data in Fig.~\ref{fig_opening} (``A'') and Fig.~\ref{fig_opening_2} (``B'') were composited starting with the top row of data set 9 and 11, respectively. The code used for composting is provided in the linked repository.

\section{Supplementary data}

\begin{figure*}[tbh]
	\includegraphics[scale=1]{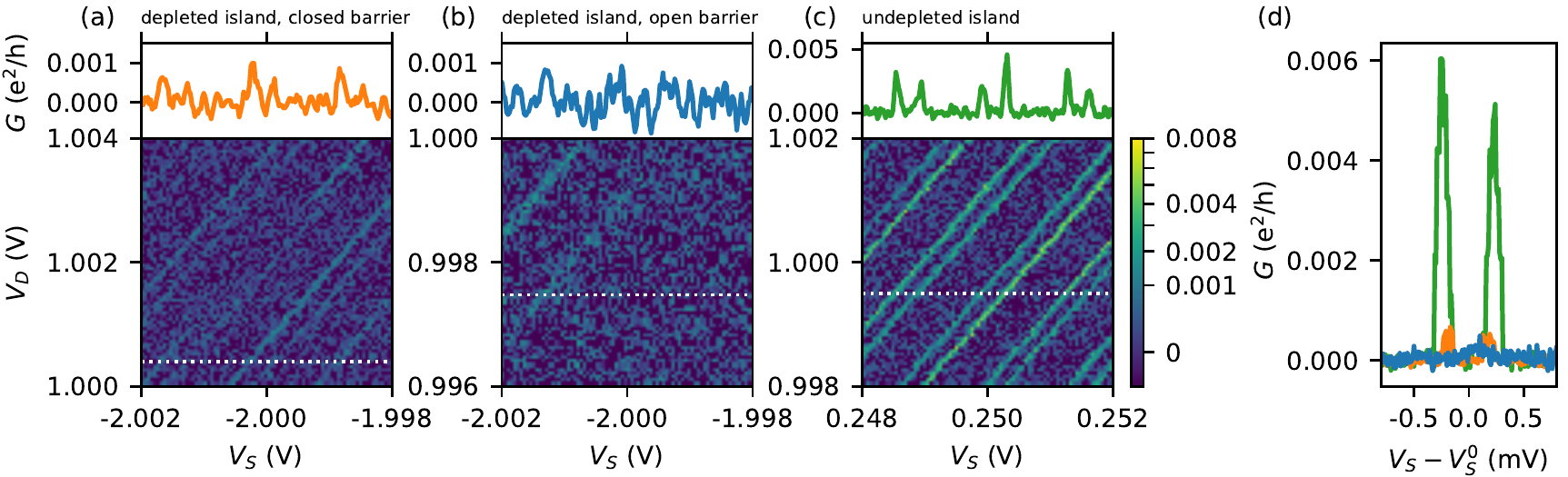}
	\caption{Effective conductance a floating dot-island in three regimes, corresponding to the $\C$ measurements presented in Fig.~\ref{fig_subgap}. The three regimes in (a-c) are: (a) depleted semiconductor under aluminum shell and small tunnel coupling; (b) depleted semiconductor and large tunnel coupling; (c) undepleted semiconductor under the aluminum shell. Shared power-law normalization of color maps was chosen to enable direct comparison of data sets.  Top panels show the cut through the data and use 3-point moving average to smooth the data. (d) Zoom in at an individual charge transition, or pair of charge transitions in the three regimes. The colors used for plotting the data correspond to the colors used in the top panels of (a-c).
	}
	\label{fig_subgap_G}
\end{figure*}

Fig.~\ref{fig_subgap_G} presents extracted effective conductance of the dot island in the three regimes: depleted island and closed barrier, depleted island and open barrier, and undepleted island. These data sets correspond, panel-by-panel, to $\C$ presented in Fig.~\ref{fig_subgap}.

\begin{figure}[tbh]
	\includegraphics[scale=1]{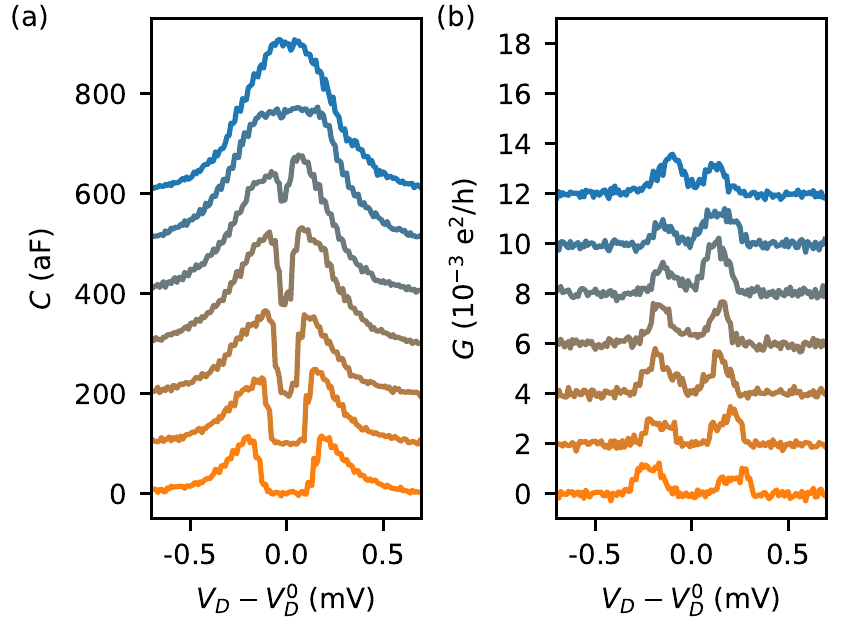}
	\caption{Data equivalent to Fig.~\ref{fig_opening}. The raw data included a complete frequency sweep at every gate setting, resulting in lower noise in the extracted value of $\G$, at the cost of significantly increased measurement time.
	}
	\label{fig_opening_freq_sweep}
\end{figure}

Fig.~\ref{fig_opening_freq_sweep} presents a data analogous to cuts in Fig.~\ref{fig_subgap}, except capacitance and conductance is extracted from full fit of the reflection coefficient as a function of frequency. This provides better signal-to-noise ratio, but does not allow to track an individual charge transition pairs due to device instabilities throughout the longer measurement time.

\begin{figure*}[tbh]
	\includegraphics[scale=1]{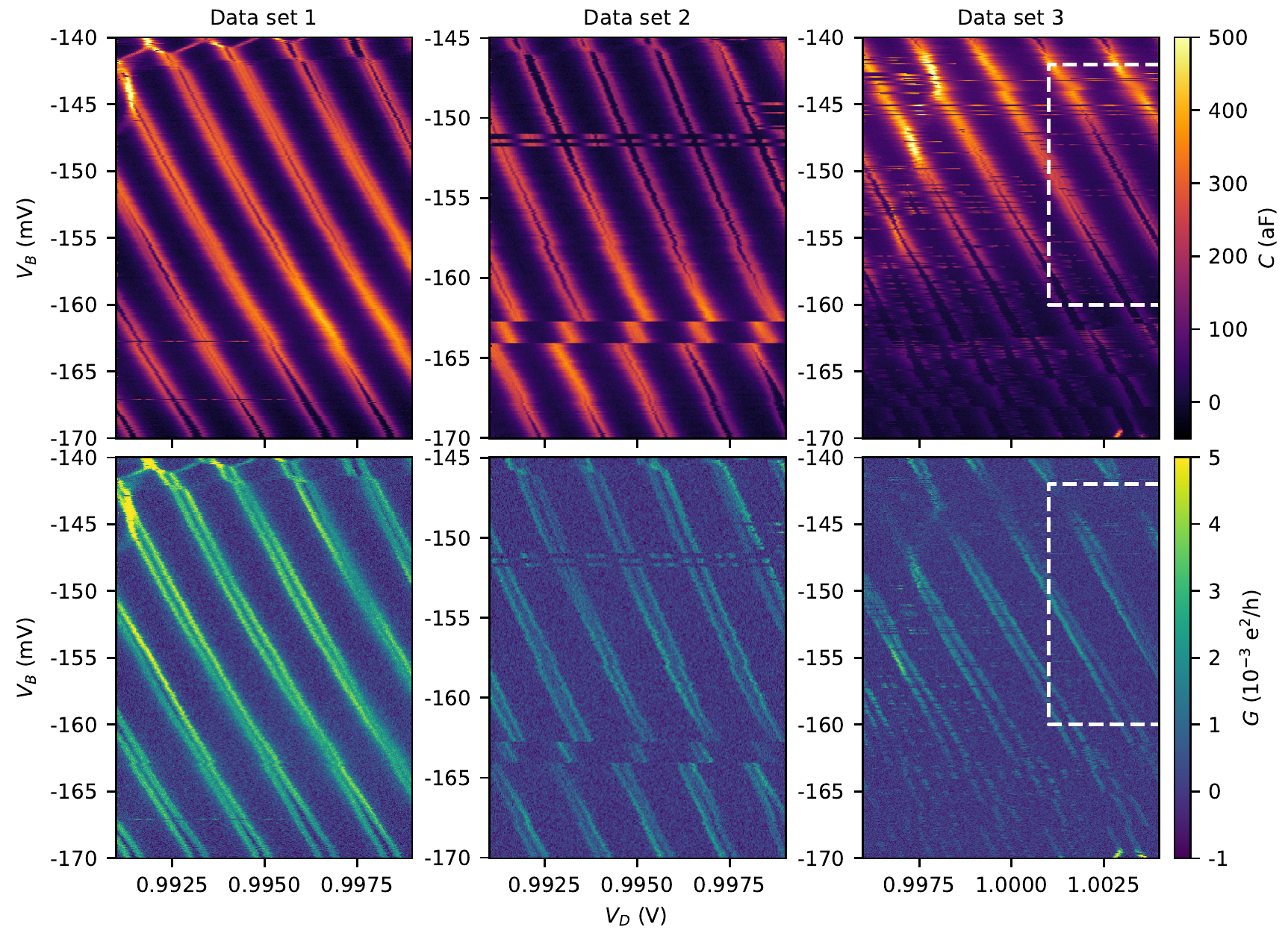}
	\caption{Additional maps of the measured effective capacitance and conductance, as a function of the barrier gate voltage $\Vb$. 
	We note the persistant double-peak strure of the charge transitions visible in the conductance data. In particular, in Data set 3, as the capacitance peaks merge, the conductance double-peak structure persists and gradually vanishes, but the peaks do not merge.
	}
	\label{fig_more_maps}
\end{figure*}

Fig.~\ref{fig_more_maps} presents additional charge stability diagrams as a function of dot plunger gate voltage $\Vd$ and barrier gate voltage $\Vb$. While the measurement ranges in all data sets are similar, the data sets represent somewhat different tunings of the dot-island, due to device instability and hysteresis. The exception is the Data set 3 and data presented in \ref{fig_opening}, which were measured immediately after. Dashed line in Data set 3 outlines edges of the range of the data in \ref{fig_opening}(a,e).

\begin{figure*}[tbh]
	\includegraphics[scale=1]{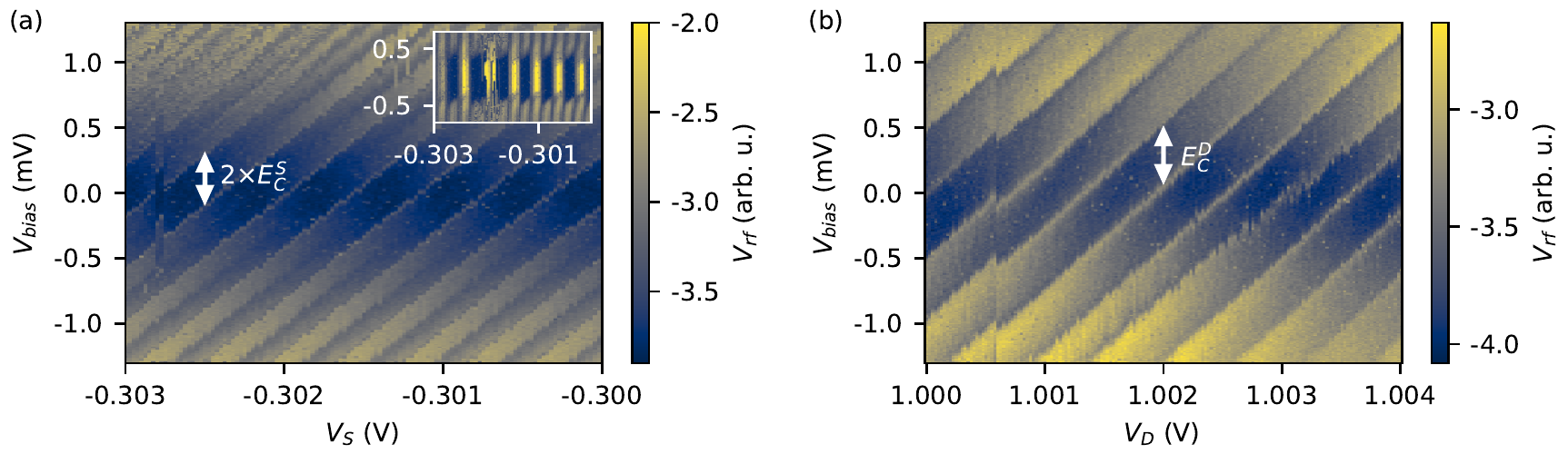}
	\caption{Coulomb diamonds of the (a) superconducting island and (b) semiconducting dot. The color maps present one of the quadratures of the reflected rf signal. The impurity charging energy $U = E_C^D$ indicated in panel (b). Since in the Coulomb diamonds measurement we can clearly distinguish only one of the slopes, it is impossible to explicitly extract the lever arm $\alpha$. Assuming the other slope to be vertical, we extract a lever arm of $\alpha = 0.82$, somewhat smaller then the value of $\alpha=0.87$ extracted from the fit in Fig.~\ref{fig_opening}(c).
	}
	\label{fig_diamonds}
\end{figure*}

Fig.~\ref{fig_diamonds} presents Coulomb diamonds of the SC island (a) and semiconducting dot (b). Due to poor galvanic contact of the nanowire to one of the metallic leads there was no measurable DC transport through the device, and the Coulomb diamonds were measured using rf-conductance with an additional spiral inductor resonator ($L=730$~nH, $f_0 \approx 313$~MHz) attached to the lead on the right side of the QD, which presumably couples through the weak galvanic link thanks to relatively large capacitance. This leads to very asymmetric Coulomb diamonds, in which we associate the vertical spacing between the linear features with (twice) a charging energy. In a separate measurement (inset) we verify that the periodicity of the Coulomb diamonds in (a) corresponds to 2e charging of the island. Extracted charging energies (defined as $e^2/C_{\Sigma}^{D/S}$) are $E_C^D = 0.50$~meV and $E_C^S = 0.20$~meV.

\bibliography{biblio}

\end{document}